\documentclass{amsart}
\usepackage{graphicx}
\usepackage{amssymb}
\usepackage{amsthm}
\usepackage{amsmath}

\newtheorem*{prop*}{Proposition}
\newtheorem*{thm*}{Theorem}

\begin {document}

\title[Wave model of the regular Sturm-Liouville operator]{Wave model of the regular 
\\Sturm-Liouville operator}
\author{Sergey Simonov}
\address{St. Petersburg Department of V. A. Steklov Institute of Mathematics of the Russian Academy of Sciences;
St. Petersburg State University;
St. Petersburg State Technological Institute (Technical University), Russia}
\email{sergey.a.simonov@gmail.com}
\maketitle

\begin{abstract}
We describe the wave functional model for the minimal (symmetric) Sturm-Liouville operator on the finite interval. We construct the wave spectrum of this operator, then, following the abstract scheme, we construct the model space of functions on the wave spectrum and introduce in that space the model operator. The latter is a matrix Sturm-Liouville operator which is unitarily equivalent to the original.
\end{abstract}

\section{Introduction}
This paper is concerned with the wave functional model of a symmetric positive definite operator. We  consider an example of the symmetric restriction of the regular Sturm-Liouville operator and build the wave model for this operator.

The wave functional model is based on the notion of the {\it wave spectrum} introduced in \cite{JOT}, see also \cite{BD_2}. In \cite{BelSim} an abstract scheme of the functional model for a symmetric positive definite operator $L_0$ was proposed. It was realized in the same work for the symmetric restriction of the Schr\"odinger operator on the half-line with a smooth potential satisfying several conditions. In that case the defect indices of the symmetry were $(1,1)$, in the present paper we consider the case of defect indices $(2,2)$. In the forthcoming paper \cite{Sim2018} a detailed construction will be developed, the purpose of this text is to give a short description of it and to formulate the main results.

\section{Abstract scheme}
We consider a symmetric operator $L_0\subset L_0^*$ acting in the Hilbert space $\mathcal H$ on the domain ${\rm Dom\,}L_0$, positive definite, $(L_0 u,u)\geqslant \varkappa \|u\|^2$, $u\in{\rm Dom\,}L_0$, with some $\varkappa>0$. $L$ denotes the Friedrichs extension of the operator $L_0$.

The abstract construction of the wave spectrum and the wave functional model was given in detail in \cite{BelSim}, \cite{Sim2018}. Due to the lack of space here we provide the necessary definitions only.

For the symmetry $L_0$ let us choose the following {\it Green system} \cite{Ryzh} (a boundary triple, \cite{MMM}): ${\mathfrak G}_{L_0}=\{{\mathcal H}, {\mathcal K};\,L_0^*, \Gamma_1, \Gamma_2\}$, where ${\mathcal K}={\rm Ker\,}L_0^*$, $\Gamma_1=L^{-1}L_0^*-I$, $\Gamma_2=P_{\mathcal K} L_0^*$ ($P_{\mathcal K}$ denotes the orthogonal projection operator on the subspace $\mathcal K$), which corresponds to the Vishik's decomposition \cite{Vishik} for the operator $L_0$.

Using $\Gamma_1$, define the following dynamical system with boundary control,
$\alpha_{L_0}$:
\begin{align}
    & u''+L_0^*u = 0,  && t>0,\label{wave eq}
    \\
    & u|_{t=0}=u'|_{t=0}=0, &&
    \\
    & \Gamma_1 u = h\,, && t\geqslant 0,
\end{align}
where $h\in C^{\infty}\left([0,\infty); {\mathcal K}\right)$ : ${\rm supp\,}h \subset (0,\infty)$. Solution $u^h$ of this system with the control $h$ is given by the formula
\begin{equation*}
    \label{weak solution u^f}
    u^h(t)=-h(t)+\int_0^t L^{-\frac12}\sin\left((t-s)L^{\frac12}\right)h''(s)ds,
\end{equation*}
these solutions are called {\it smooth waves}. The sets
\begin{equation}\label{U^t}
    {\mathcal U}^t_{L_0}=\{u^h(t),\ h\in C^{\infty}\left([0,t]; {\mathcal K}\right):{\rm supp\,}h \subset (0,t]\}
\end{equation}
are called {\it reachable sets} for the time $t\geqslant0$, and
\begin{equation*}\label{U}
    {\mathcal U}_{L_0}=\bigcup_{t>0}{\mathcal U}^t_{L_0}
\end{equation*}
is the total reachable set, the linear set of smooth waves. If the operator $L_0$ is completely non-selfadjoint, then $\overline{{\mathcal U}}_{L_0}={\mathcal H}$, see \cite{BD_2}.

On the lattice $\mathfrak L(\mathcal H)$ of subspaces \cite{Birkhoff} of the Hilbert space $\mathcal H$ we introduce an isotony (see \cite{BelSim}) related to the operator $L$. Consider the system
\begin{align*}
    & v''+L v = g\,,  && t>0,
    \\
    & v|_{t=0}=v'|_{t=0}=0\,,
\end{align*}
with $g\in C^{\infty}\left([0,\infty); {\mathcal H}\right):{\rm supp\,}g \subset (0,\infty)$. Solution of this system is given by the formula
\begin{equation*}
    v^g(t)=\int_0^tL^{-\frac{1}{2}}\sin\left((t-s)L^{\frac{1}{2}}\right)g(s)ds.
\end{equation*}
The mapping on ${\mathfrak L}(\mathcal H)$
\begin{equation*}
I_{L}^t(\mathcal G)=\overline{\left\{v^g(t),g\in C^{\infty}([0,t];\mathcal G),{\rm supp\,}g,
\subset (0,t]\right\}},
\end{equation*}
$\mathcal G\in{\mathfrak L}(\mathcal H)$, (while $I_{L}^0(\mathcal G)=\mathcal G$) is an isotony of the lattice ${\mathfrak L}(\mathcal H)$ (see \cite{JOT}, \cite{BelSim}) and is called the {\it wave isotony}.

There exists the minimal sublattice $\mathfrak L_{L_0}$ of the lattice $\mathfrak L(\mathcal H)$ which contains all the subspaces $\overline{\mathcal U^t_{L_0}}$, $t>0$, and is invariant under the wave isotony $I_L^t$. Consider the set of $\mathfrak L(\mathcal H)$-valued functions $I_L\mathfrak L_{L_0}=\{I_L^t(\mathcal G), \mathcal G\in\mathfrak L_{L_0}\}$ and its closure in the sense of the point-wise convergence in the lattice $\mathcal F(\mathcal H)$ of functions from $[0,\infty)$ to $\mathfrak L(\mathcal H)$ (see \cite{BelSim}). Atoms of this closure form a set which is called the wave spectrum of the operator $L_0$,
\begin{equation*}
\Omega_{L_0}={\rm At}(\overline{I_L\mathfrak L_{L_0}}).
\end{equation*}
One can define a topology, a metric (under additional assumptions) on the wave spectrum, the boundary $\partial\Omega_{L_0}$ of the wave spectrum.

It will be convenient to introduce further constructions of the wave model on an example, so we pass directly to it.

\section{Regular Sturm-Liouville operator}
We consider the operator $L_0$ in ${\mathcal H} = L_2(0,l)$ (with $l\in(0,\infty)$) on the domain
$$
{\rm Dom\,} L_0=\{u\in H^2(0,l):
u(0)=u'(0)=u(l)=u'(l)=0\},
$$
which acts by the rule
\begin{equation*}
L_0u=-u''+qu,
\end{equation*}
where $q\in C^{\infty}[0,l]$ is a smooth function such that the operator $L_0$ is positive definite. It has defect indices $(2,2)$. Then the operator $L$, the Friedrichs extension of the operator $L_0$, is the self-adjoint Sturm-Liouville operator with the Dirichlet boundary conditions at both ends of the interval.

\subsection*{The dynamical system with boundary control}
From the Green system we only need the subspace ${\mathcal K}={\rm Ker\,}L_0^*=\{u:-u''+qu=0\}$ and
the first boundary operator
\begin{equation*}\label{Gamma-1}
\Gamma_1u=-\frac{u(l)}{\phi_0(l)}\phi_0-\frac{u(0)}{\phi_l(0)}\phi_l,
\end{equation*}
where $\phi_0$ and $\phi_l$ are solutions of the equation $-u''+qu=0$ with the initial conditions $\phi_0(0)$, $\phi_0'(0)=1$, and $\phi_l(l)=0$, $\phi_l'(l)=1$.

The system $\alpha_{L_0}$ can be written as
\begin{align*}
& u_{tt}-u_{xx}+qu = 0,  && x\in(0,l),\,t>0,
\\
& u|_{t=0}=u_t|_{t=0}=0, && x\in[0,l],
\\
& u|_{x=0} = f_0(t), && t\geqslant 0,
\\
& u|_{x=l} = f_l(t), && t\geqslant 0,
\end{align*}
with the control consisting of two functions $f_0,f_l\in C^{\infty}[0,\infty)$: ${\rm supp\,}f_0,{\rm supp\,}f_l \subset (0,\infty)$. Reachable sets of this system have the form
\begin{equation*}\label{UT SL}
\mathcal U_{L_0}^t=
\left\{
\begin{array}{ll}
\{u\in C^{\infty}[0,l]:{\rm supp}\, u\subset[0,t)\cup(l-t,l]\},&t\leqslant \frac l2,
\\
C^{\infty}[0,l],&t>\frac l2.
\end{array}
\right.
\end{equation*}

\subsection*{The wave isotony and the wave spectrum}
The wave isotony acts on subspaces of the form $L_2(a,b)$, $0<a<b<l$, as
\begin{equation*}\label{isotony on interval}
I^t_L(L_2(a,b))=L_2((a,b)^t),
\end{equation*}
where $(a,b)^t$ is the metric neighborhood of the interval $(a,b)$ in $(0,l)$. For a set $E$ consisting of a finite number of non-intersecting intervals one also has
\begin{equation*}
I_L^t(L_2(E))=L_2(E^t).
\end{equation*}
It follows that the sublattice $\mathfrak L_{L_0}$ consists of subspaces of square integrable functions supported by sets, which are finite unions of non-intersecting intervals and are symmetric with respect to the point $\frac l2$. One can prove that the closure in $\mathcal F(\mathcal H)$ of the set $I_L\mathfrak L_{L_0}$, functions ``growing'' with time from such subspaces, consists of subspaces of square integrable functions supported by (at the moment $t$) metric neighborhoods of the distance $t$ of subsets of $[0,l]$, which are symmetric with respect to $\frac l2$ and consist of a countable number of non-intersecting intervals as well as isolated points. The wave spectrum is the set of atoms of this closure, it consists of the elements of the form
\begin{equation*}
\omega_x:t\mapsto L_2\left((\{x\}\cup\{l-x\})^t\right).
\end{equation*}

\noindent
{\bf Proposition 1} (\cite{Sim2018}).
{\it
\begin{equation*}
\Omega_{L_0}=\left\{\omega_x,\ x\in\left[0,\frac l2\right]\right\}.
\end{equation*}
}

The wave spectrum is therefore naturally bijective to $[0,\frac l2]$, a half of the original interval $[0,l]$.

Each atom $\omega\in\Omega_{L_0}$ generates a resolution of identity
\begin{equation*}
E_{\omega}(t)=
\left\{
\begin{array}{ll}
P_{\omega(t)},\ &t\geqslant0,
\\
0,&t<0,
\end{array}
\right.
\end{equation*}
which determines the eikonal operator
\begin{equation*}
\tau_{\omega}=\int_{\mathbb R}tdE_{\omega}(t).
\end{equation*}
This is an unbounded operator, however, the difference of two eikonals is bounded and one can define the function
\begin{equation*}
\tau(\omega_1,\omega_2)=\|\tau_{\omega_1}-\tau_{\omega_2}\|
=|x_{\omega_1}-x_{\omega_2}|
\end{equation*}
which can serve as a metric on $\Omega_{L_0}$ (here $x_{\omega}$ is the point such that $\omega=\omega_{x_{\omega}}$). The ``boundary'' of $\Omega_{L_0}$ consists of one element $\omega_0$ (see \cite{Sim2018}).

\subsection*{The wave model}
It is possible to choose a function $e\in\mathcal K$ such that for every $\omega\in\Omega_{L_0}$ one has
$|e(x_{\omega})|^2+|e(l-x_{\omega})|^2\neq0$. Then
\begin{equation*}
\frac{\|P_{\omega(t)}u\|^2}{\|P_{\omega(t)}e\|^2}\underset{t\to+0}
\longrightarrow
\frac{|u(x_{\omega})|^2+|u(l-x_{\omega})|^2}{|e(x_{\omega})|^2+|e(l-x_{\omega})|^2},
\end{equation*}
and one can define a sesquilinear form on smooth waves
\begin{equation*}
\langle u,v\rangle_{\omega}
=\frac{u(x_{\omega})\overline{v(x_{\omega})}+u(l-x_{\omega})\overline{v(l-x_{\omega})}}{|e(x_{\omega})|^2+|e(l-x_{\omega})|^2},\ \end{equation*}
$u,v\in\mathcal U_{L_0}$. Factorize the linear set $\mathcal U_{L_0}$ by the null subspace of this form:
\begin{equation*}
u\underset{\omega}\sim v\Leftrightarrow
\left\{
\begin{array}{l}
u(x_{\omega})=v(x_{\omega}),
\\
u(l-x_{\omega})=v(l-x_{\omega}).
\end{array}
\right.
\end{equation*}
The resulting subspace has dimension $2$, denote it by $\mathcal U_{L_0,\omega}^{\rm w}$. This is the space of values at the point $\omega\in\Omega_{L_0}$. The following equality holds:
\begin{multline*}
(u,v)_{\mathcal H}=\int_0^{\frac l2}(u(x)\overline{v(x)}+u(l-x)\overline{v(l-x)})dx
\\
=\int_0^{\frac l2}\langle[u](\omega_x),[v](\omega_x)\rangle_{\mathcal U_{L_0,\omega_x}^{\rm w}}\rho(x)dx
=\int_{\Omega_{L_0}}\langle[u](\omega),[v](\omega)\rangle_{\mathcal U_{L_0,\omega}^{\rm w}}d\mu(\omega),
\end{multline*}
where $[u],[v]$ are equivalence classes,
\begin{equation*}
\rho(x)=|e(x)|^2+|e(l-x)|^2,
\end{equation*}
$\mu$ is the image of the measure $\rho(x)dx$ on $\Omega_{L_0}$ under the map $x\mapsto\omega_x$. The space
\begin{equation*}
\mathcal H^{\rm w}=\oplus\int_{\Omega_{L_0}}\mathcal U_{L_0,\omega}^{\rm w}d\mu(\omega)
\end{equation*}
is the model space of values, the unitary operator $W^{\rm w}$ acts from $\mathcal H$ to $\mathcal H^{\rm w}$ so that $W^{\rm w}u=[u](\cdot)$ for $u\in\mathcal U_{L_0}$. In $\mathcal H^{\rm w}$ acts the operator ${L_0^{\rm w}}^*=W^{\rm w}L_0^*{W^{\rm w}}^*$, and we can obtain its graph with the boundary control, using \eqref{wave eq}, \eqref{U^t}, and the fact that $(u^h)'=u^{(h')}$ for $h\in C^{\infty}([0,t];\mathcal K)$ with ${\rm supp\,}h\subset(0,\infty)$:
\begin{multline*}
{\rm Graph\,}(W^{\rm w}L_0^*|_{\mathcal U_{L_0}}{W^{\rm w}}^*)
=\{(W^{\rm w}u,W^{\rm w}L_0^*u),u\in\mathcal U_{L_0}\}
\\
=\{(W^{\rm w}u^h(t),-W^{\rm w}u^{(h'')}(t)),\
h\in C^{\infty}([0,t];\mathcal K),{\rm supp\,}h\subset(0,t],t\geqslant0\}.
\end{multline*}
In our case $\overline{L_0^*|_{\mathcal U_{L_0}}}=L_0^*$, therefore the graph of ${L_0^{\rm w}}^*$ is the closure of the obtained graph.

\subsection*{The coordinate representation}
The coordinate wave model can be also constructed, in the following way. Pick two linearly independent vectors $e_1,e_2\in\mathcal K$. In every space $\mathcal U_{L_0,\omega}^{\rm w}$ the vectors $[e_1](\omega)$ and $[e_2](\omega)$ form a base. Let us take
\begin{equation*}
\hat u(x_{\omega})=
\begin{pmatrix}
  \langle u,e_1\rangle_{\omega} \\
  \langle u,e_2\rangle_{\omega} \\
\end{pmatrix}
\end{equation*}
as the ``coordinate'' value of $u\in\mathcal U_{L_0}$ on $\omega$. This lets us use $\mathbb C^2$ instead of $\mathcal U_{L_0,\omega}^{\rm w}$ in the model. It turns out that
\begin{equation*}
\langle[u](\omega),[v](\omega)\rangle_{\mathcal U_{L_0,\omega}^{\rm w}}=(G^{-1}(\omega)\hat u(x_{\omega}),\hat v(x_{\omega}))_{\mathbb C^2},
\end{equation*}
where
\begin{equation*}
G(\omega)=
\begin{pmatrix}
  \langle e_1,e_1\rangle_{\omega} & \langle e_2,e_1\rangle_{\omega} \\
  \langle e_1,e_2\rangle_{\omega} & \langle e_2,e_2\rangle_{\omega} \\
\end{pmatrix}
\end{equation*}
is the Gram matrix. Thus instead of $\mathcal H^{\rm w}$ we get
\begin{equation*}
\mathcal H^{\rm c}=L_2\left(\left(0,\frac l2\right),G^{-1}(\omega)\rho(x_{\omega})dx_{\omega}; \mathbb C^2\right).
\end{equation*}
The unitary operator $W^{\rm c}$ acts from $\mathcal H$ to $\mathcal H^{\rm c}$ so that $W^{\rm c}u=\hat u(\cdot)$ for  $u\in\mathcal U_{L_0}$, the graph of the operator
\begin{equation*}
{L_0^{\rm c}}^*=W^{\rm c}L_0^*{W^{\rm c}}^*
\end{equation*}
is constructed analogously to the graph of the operator ${L_0^{\rm w}}^*$. This completes the construction of the wave functional model, and the operator ${L_0^{\rm c}}^*$ turns out to be a differential operator of the second order in $L_2\left(\left(0,\frac l2\right),G^{-1}(\omega)\rho(x_{\omega})dx_{\omega}; \mathbb C^2\right)$.

\noindent
{\bf Proposition 2} (\cite{Sim2018}).
{\it The operator ${L_0^{\rm c}}^*$ acts by the rule
\begin{equation*}
({L_0^{\rm c}}^*\hat u)(x)=-\hat u''(x)+\hat P(x)\hat u'(x)+\hat Q(x)\hat u(x),
\end{equation*}
where
\begin{equation*}\label{P-hat}
\hat P(x)=-2T(x){T^{-1}}'(x),
\end{equation*}
\begin{equation*}\label{Q-hat}
\hat Q(x)=T(x)Q(x)T^{-1}(x)-T(x){T^{-1}}''(x),
\end{equation*}
and
\begin{equation*}
Q(x)=
\begin{pmatrix}
  q(x) & 0 \\
  0 & q(l-x) \\
\end{pmatrix},
\end{equation*}
\begin{equation*}\label{T}
T(x)=\frac1{\rho(x)}
\begin{pmatrix}
  \overline{e_1(x)} & \overline{e_1(l-x)} \\
  \overline{e_2(x)} & \overline{e_2(l-x)} \\
\end{pmatrix}.
\end{equation*}
}

Apparently, the model operator looks similar to the original $L_0^*$. An observer is able to construct the coordinate functional wave model and to recover the potential $q$ from the matrix coefficients $\hat P$ and $\hat Q$ up to reflection with respect the point $\frac l2$.

\section*{Acknowledgements}
The work was supported by the grants RFBR 16-01-00443À and RFBR 16-01-00635À.

\end {document}